\documentclass[iop]{emulateapj}

\usepackage{natbib}
\usepackage{amsmath}
\usepackage{gensymb}
\usepackage{chngcntr}

\shorttitle{Discovery of an extremely wide-angle bipolar outflow in AFGL 5142} \shortauthors{Liu et al.}

\begin{document}

\title{Discovery of an extremely wide-angle bipolar outflow in AFGL 5142 }
\author{Tie Liu\altaffilmark{1}, Qizhou Zhang\altaffilmark{2}, Kee-Tae Kim\altaffilmark{1}, Yuefang Wu\altaffilmark{3}, Chang-Won Lee\altaffilmark{1,4}, Paul F. Goldsmith\altaffilmark{5}, Di Li\altaffilmark{6,7}, Sheng-Yuan Liu\altaffilmark{8}, Huei-Ru Chen\altaffilmark{8,9}, Ken'ichi Tatematsu\altaffilmark{10}, Ke Wang\altaffilmark{11}, Jeong-Eun Lee\altaffilmark{12}, Sheng-Li Qin\altaffilmark{13}, Diego Mardones\altaffilmark{14}, Se-Hyung Cho\altaffilmark{1}   }
\altaffiltext{1}{Korea Astronomy and Space Science Institute 776, Daedeokdae-ro, Yuseong-gu, Daejeon, Republic of Korea 305-348; liutiepku@gmail.com}
\altaffiltext{2}{Harvard-Smithsonian Center for Astrophysics, 60 Garden Street, Cambridge, MA 02138, USA}
\altaffiltext{3}{Department of Astronomy, Peking University, 100871, Beijing China}
\altaffiltext{4}{University of Science \& Technology, 217 Gajungro, Yuseong-gu, 305-333 Daejeon, Korea}
\altaffiltext{5}{Jet Propulsion Laboratory, California Institute of Technology, 4800 Oak Grove Drive, Pasadena, CA 91109, USA}
\altaffiltext{6}{ Key Laboratory for Radio Astronomy, National Astronomical Observatories, Chinese Academy of Science, A20 Datun Road, Chaoyang District, Beijing 100012, China}
\altaffiltext{7}{Key Laboratory for Radio Astronomy, Chinese Academy of Sciences, Nanjing 210008, China}
\altaffiltext{8}{Institute of Astronomy and Astrophysics, Academia Sinica, Taipei, Taiwan}
\altaffiltext{9}{Institute of Astronomy and Department of Physics, National Tsing Hua University, Hsinchu, Taiwan}
\altaffiltext{10}{National Astronomical Observatory of Japan, 2-21-1 Osawa, Mitaka, Tokyo 181-8588, Japan}
\altaffiltext{11}{European Southern Observatory, Karl-Schwarzschild-Str. 2, D-85748 Garching bei M\"{u}nchen, Germany}
\altaffiltext{12}{School of Space Research, Kyung Hee University, Yongin-Si, Gyeonggi-Do 446-701, Korea}
\altaffiltext{13}{Department of Astronomy, Yunnan University, and Key Laboratory of Astroparticle Physics of Yunnan Province, Kunming, 650091, China}
\altaffiltext{14}{Departamento de Astronom\'{\i}a, Universidad de Chile, Casilla 36-D, Santiago, Chile}

\begin{abstract}
Most bipolar outflows are associated with individual young stellar objects and have small opening angles.
Here we report the discovery of an extremely wide-angle ($\sim$180$\arcdeg$) bipolar outflow (``EWBO") in a cluster forming region AFGL 5142 from low-velocity emission of the HCN (3-2) and HCO$^{+}$ (3-2) lines. This bipolar outflow is along a north-west to south-east direction with a line-of-sight flow velocity of about 3 km~s$^{-1}$ and is spatially connected to the high-velocity jet-like outflows. It seems to be a collection of low-velocity material entrained by the high-velocity outflows due to momentum feedback. The total ejected mass and mass loss rate due to both high velocity jet-like outflows and the ``EWBO" are $\sim$24.5 M$_{\sun}$ and $\sim1.7\times10^{-3}$ M$_{\sun}$~yr$^{-1}$, respectively. Global collapse of the clump is revealed by the ``blue profile" in the HCO$^{+}$ (1-0) line. A hierarchical network of filaments was identified in NH$_{3}$ (1,1) emission. Clear velocity gradients of the order of 10 km~s$^{-1}$~pc$^{-1}$ are found along filaments, indicating gas inflow along the filaments. The sum of the accretion rate along filaments and mass infall rate along the line of sight is $\sim$3.1$\times10^{-3}$ M$_{\sun}$~yr$^{-1}$, which exceeds the total mass loss rate, indicating that the central cluster is probably still gaining mass. The central cluster is highly fragmented and 22 condensations are identified in 1.1 mm continuum emission. The fragmentation process seems to be determined by thermal pressure and turbulence. The magnetic field may not play an important role in fragmentation.
\end{abstract}


\section{Introduction}

Outflows in high-mass star-forming regions have been found with a
high detection rate in both single-dish surveys
\citep{wu04,zhang05,kim06,qin08a} and high resolution interferometer
studies \citep{su04,liu11a,liu11b,liu13a,liu15,qin08b,qin08c,qiu07,qiu09,qiu11a,qiu11b,zhang07,ren11,zhu11,wang11}. Very collimated jet-like outflows are frequently found toward young \citep{wang11} or isolated \citep{ren11,zhu11} high-mass star forming regions. In evolved cluster forming regions \citep{qiu09,qiu11b,liu11b}, outflows often show complicated and less collimated structures.

Located at a distance of 1.8 kpc \citep{sne88}, AFGL 5142 is a high-mass star forming region containing a cluster of continuum cores \citep{zhang07,pal11,pal13}. The CO and SO line emission from interferometric observations showed at least three well collimated molecular outflows originating from the center of
the dust core \citep{zhang07,pal11,pal13}. Observations of the H$_{2}$O maser (and radio continuum) emission revealed a collimated bipolar molecular outflow (and ionized jet) from the brightest continuum core ``MM1", which has an infalling envelope (radius of 300 AU and infall velocity of 5 km~s$^{-1}$) \citep{God11}. Thus MM1 was identified as the driving source of the large-scale CO/SO outflows \citep{God11}. The outflows are heating the surroundings as implied by the enhanced temperatures seen in the ratio maps of different transitions of NH$_{3}$ emission \citep{zhang02}. In order to further explore the outflow properties and outflow feedback in AFGL 5142, we have carried out interferometric observations of this region with the Submillimeter Array (SMA). In this paper, we report the discovery of an extremely wide-angle ($\sim$180$\arcdeg$) bipolar outflow in AFGL 5142 from low-velocity emission in the HCN (3-2) and HCO$^{+}$ (3-2) lines. This outflow is entirely different from those well-defined and collimated bipolar outflows associated with young stellar objects or other disordered outflows in cluster forming regions. It seems to be a collection of low-velocity materials entrained by high-velocity jet-like outflows.

\section{Observations}

The observations of AFGL 5142 were carried out with the SMA in
November 2011 in its extended configuration and in January 2012 in its compact configuration. The phase reference center was R.A.(J2000)~=~05$^{\rm h}$30$^{\rm m}$48.02$^{\rm s}$ and
DEC.(J2000)~=~$33\arcdeg47\arcmin54.48\arcsec$. In both observations, QSOs 3c111 and 0510+180 were observed for gain correction and Callisto was used for flux-density calibration. The absolute flux level is accurate
to about 15\%. Bandpass was corrected by observing QSOs 3c454.3 in its extended configuration and 3c279 in its compact configuration, respectively. The 345 GHz receivers were tuned to 265 GHz for the lower sideband (LSB)
and 275~GHz for the upper sideband (USB).  The frequency spacing across the spectral band is 0.812~MHz or $\sim$0.9 km s$^{-1}$. The visibility data
were calibrated with the IDL superset MIR package and imaged with MIRIAD and CASA packages. The 1.1 mm continuum data were acquired by averaging all the line-free channels over both the upper and lower spectral bands. The MIRIAD task ``selfcal" was employed to perform self-calibration on the continuum data. The gain solutions from the self-calibration were applied to the line data. The synthesized beam size and 1 $ \sigma$ rms of the continuum emission from combining both compact and extended configuration data are $1.41\arcsec\times1.14\arcsec$ (PA=65.8$\arcdeg$) and $\sim$1.5 mJy~beam$^{-1}$, respectively. We smoothed the spectral lines to a spectral resolution of 1 km~s$^{-1}$ and the corresponding 1$ \sigma$ rms for lines is $\sim$80 mJy~beam$^{-1}$ per channel. The synthesized beam size and 1 $ \sigma$ rms of the continuum emission from extended configuration data only are $0.84\arcsec\times0.62\arcsec$ (PA=87.6$\arcdeg$) and $\sim$2 mJy~beam$^{-1}$, respectively.

To recover the missing short spacing information in the HCN (3-2) line, we used the single-dish data obtained from the CSO archive. The observations were carried out in January 2004. The spectral resolution was 0.054 km~s$^{-1}$. The weather was good during observations with $\tau\sim0.1$ and system temperature of 490 K. The beam size and beam efficiency of the CSO data are $\sim$28$\arcsec$ and 0.67, respectively. The map size is $1.5\arcmin\times1.5\arcmin$ with a spacing of 10$\arcsec$. The MIRIAD task ``mosmem" was used to combine the SMA and the CSO data.

We observed AFGL 5142 in the J=1-0 transitions of HCO$^{+}$ and H$^{13}$CO$^{+}$ as well as the J=14-13 and 15-14 transitions of HC$_{3}$N with the Korean VLBI Network (KVN) 21 m telescope at Yonsei station in single dish mode \citep{kim11}. The observations were carried out in November 2014. The spectral resolution and system temperature for J=1-0 of HCO$^{+}$ and H$^{13}$CO$^{+}$ are $\sim$0.052 km~s$^{-1}$ and 230 K ($\tau\sim$0.12), respectively. The spectral resolution and system temperature for HC$_{3}$N (14-13) and (15-14) are $\sim$0.034 km~s$^{-1}$ and 260 K ($\tau\sim$0.37), respectively. We did single-pointing observations with on source time of about 15 minutes to achieve an rms level of T$_{A}^{*}<$0.1 K. The mean-beam efficiencies of the telescope are 0.41 and 0.40 at 89 and 136 GHz, respectively.

The 450 $\micron$ and 850 $\micron$ continuum data were obtained from JCMT archive. The JCMT/SCUBA observations were conducted in October 1997. The beam sizes of SCUBA at 450 $\micron$ and 850 $\micron$ are 9.8$\arcsec$ and 14.4$\arcsec$, respectively. We also obtained publicly available Herschel data from Herschel archive. The proposal ID of Herschel observations is ``OT1\_smolinar\_5". The 70 and 160 $\micron$ continuum data were obtained with the Photodetector Array Camera \& Spectrometer \citep[PACS,][]{pog10}. The 250, 350 and 500 $\micron$ continua were observed with the
Spectral and Photometric Imaging Receiver \citep[SPIRE,][]{gri10}. The angular resolutions at 70, 160, 250, 350 and 500 $\micron$ wavelengths are about 6$\arcsec$, 12$\arcsec$, 18$\arcsec$, 25$\arcsec$, and 36$\arcsec$, respectively. We covert the units of Herschel maps to Jy/pixel for aperture photometry.

We also used NH$_{3}$ (1,1) data for analysis. Observations of NH$_{3}$ (1,1) line were conducted using the VLA on 1999 March 27, 2000 April 4 and 2000 September 3 and 4 in the compact D array, C array and D array, respectively. The combined data have a beam size of $3.58\arcsec\times2.64\arcsec$, with P.A.=-14$\arcdeg$. The spectral resolution of NH$_{3}$ (1,1) line is $\sim$0.6 km~s$^{-1}$. The details of NH$_{3}$ observations are presented in \cite{zhang02}.

\section{Results}

\subsection{Continuum emission}

\subsubsection{Herschel and SCUBA continuum emission}

\begin{figure}
\centering
\includegraphics[angle=0,scale=0.4]{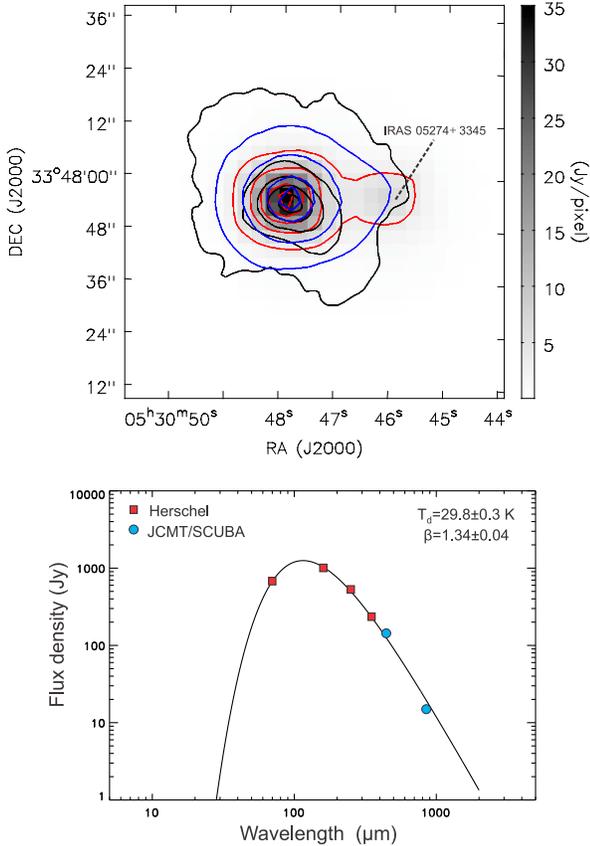}
\caption{Upper panel: The 70 $\micron$ continuum emission is shown as the red contours and gray-scale image. The 160 $\micron$ and 450 $\micron$ continuum emissions are shown as blue and black contours, respectively. The contour levels are from 10\% to 90\% in steps of 20\% of the peak values. The peak values for 70 $\micron$, 160 $\micron$ and 450 $\micron$ continuum emissions are 35.1 Jy/pixel (1 pixel size=3.2$\arcsec$), 32.2 Jy/pixel (1 pixel size=3.2$\arcsec$) and 45.5 Jy/beam,respectively. Lower panel: The best-fit SED is shown as a solid line. The blue filled circles indicate the flux densities derived from JCMT/SCUBA. The red filled boxes denote flux densities derived from Herschel. The flux densities were measured with a photometric aperture radius of 20$\arcsec$.}
\end{figure}

In the upper panel of Figure 1, we present the Herschel 70 $\micron$ and 160 $\micron$ continuum emission as well as SCUBA 450 $\micron$ continuum emission. The 70 $\micron$ emission reveals two objects. The western one is IRAS 05274+3354. The eastern one (AFGL 5142), which shows much stronger emission at longer wavelengths than IRAS 05274+3354, is an active star forming clump. In this paper, we focus only on the eastern clump (AFGL 5142). In contrast to 70 $\micron$ continuum, the 160 $\micron$ and 450 $\micron$  continuum emission only show a single clump peaked at the eastern clump. The 250, 350, 500 and 850 $\micron$ continuum images, which are not shown here, also reveal a single clump as the 160 $\micron$ and 450 $\micron$  continuum emission. From a Gaussian fit, we find that the mean full width of half maximum (FWHM) of deconvolved major and minor axes of the 450 $\micron$  continuum emission are $\sim$20$\arcsec$ ($\sim$36000 AU at a distance of 1.8 kpc). We take 36000 AU as the clump radius.

We measured the aperture fluxes of AFGL 5142 with an aperture radius of 20$\arcsec$. The measured fluxes of the 70, 160, 250, 350, 450 and 850 $\micron$ continuum emission are 678.6$\pm$9, 1007.1$\pm$10.8, 530.1$\pm$8.7, 235.4$\pm$10.2, 142.8$\pm$13.6 and 15$\pm$1.3 Jy, respectively. We did not use the 500 $\micron$ continuum data because of its low angular resolution. The spectral energy distribution (SED) of AFGL 5142 is shown in the lower panel of Figure 1. The continuum
emission at the frequency $\nu$ from thermal dust subtending a solid angle $\Omega$ in a molecular clump which have a dust temperature $T_{\rm d}$
and a total mass of gas and dust $M$, can be described as \citep{wils09,zhu10}
\begin{equation}
\label{} S_{\nu} = B_{\nu}(T_{\rm
d})(1-e^{-\tau_{\nu}})\Omega
\end{equation}
where $S_{\nu}$ is the flux of the dust emission, $\tau_{\nu}$ is optical depth, and $B_{\nu}(T_d)$ is the Planck
function.

\begin{equation}
\tau_{\nu} = M\kappa_{\nu}/D^2\Omega = \mu m_{H}\kappa_{\nu}N_{H_{2}}
\end{equation}
where $\kappa_{\nu}$, D, $\mu$=2.37, $m_{H}$, and $N_{H_{2}}$ are the dust opacity per unit gas mass, distance, mean molecular weight, atomic hydrogen mass, and H$_{2}$ column density, respectively.
Assuming a gas to dust ratio of 100, the dust opacity per unit gas mass is $\kappa_{\nu}=\kappa_{0}(\frac{\nu}{\nu_{0}})^{\beta}$, where $\kappa_{0}=0.01$ cm$^{2}$g$^{-1}$ is the dust opacity at
$\nu_{0}$=230 GHz derived from \cite{oss94}.

The SED can be well fitted with equation (1) assuming a single dust component. The thermal dust
model has three free parameters ($M$, $T_{\rm d}$ and $\beta$). We used a nonlinear least-squares method (the Levenberg--Marquardt algorithm coded within IDL) to fit the observed SED. A total dust and gas mass of M=210$\pm$12 M$_{\sun}$, a dust temperature of $T_{\rm d}$=29.8$\pm$0.3 K, and a dust opacity index of $\beta$=1.34$\pm$0.04 were derived from the best fit model.

The mean particle volume density n can be calculated as:
\begin{equation}
n=\frac{M}{\frac{4}{3}\pi R^{3}\mu m_{H}}
\end{equation}

The volume density of the clump is $\sim1.6\times10^{5}$ cm$^{-3}$. The radius, mass and volume density of the clump are summarized in Table 1.

\begin{figure}
\centering
\includegraphics[angle=0,scale=0.4]{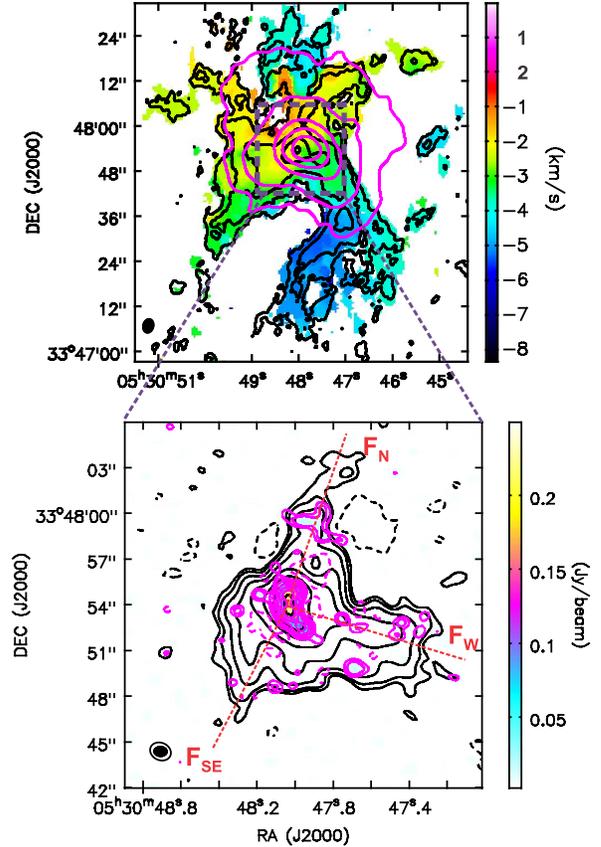}
\caption{ Upper panel: The Moment 1 map of the main hyperfine component of NH$_{3}$ (1,1) is shown as color image. The integrated intensity of the main hyperfine component of NH$_{3}$ (1,1) is shown in black contours. The contour levels are [3, 5, 10, 20, 40]$\times$1 $\sigma$. The rms uncertainty $\sigma$ is 5 mJy~beam$^{-1}$~km~s$^{-1}$. The flux intensity of the 450 $\micron$ continuum is shown in pink contours. The contour levels are 10\%, 20\%, 40\%, 60\% and 80\% of the peak value 45.5 Jy/beam. Lower panel: The 1.1 mm continuum emission from the combined SMA compact and extended array data is shown in black contours. The contours are (-3,3,5,10,20,40,80,160,320)$\times$1.5 mJy/beam (1 $\sigma$). The 1.1 mm continuum emission from the SMA extended array data is shown as color image and pink contours. The contours are (-3,3,5,10,20,40,80,160,320)$\times$2.0 mJy/beam (1 $\sigma$). The red dashed lines mark the directions of the three elongated structures. }
\end{figure}

\subsubsection{SMA 1.1 mm continuum}

\begin{figure}
\centering
\includegraphics[angle=-90,scale=0.35]{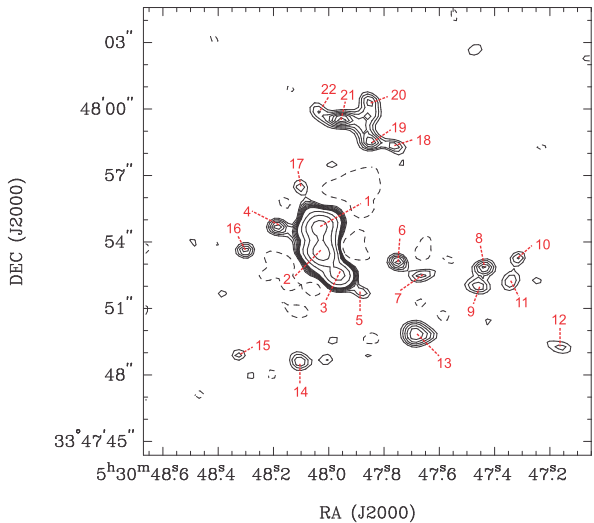}
\caption{The 1.1 mm continuum emission from the SMA extended array data is shown in black contours. The contours are (-3,3,4,5,6,7,8,9,10,20,40,80,160,320)$\times$2.0 mJy/beam (1 $\sigma$). The names of condensations are labeled with numbers.}
\end{figure}

\begin{deluxetable*}{cccccccccc}
\centering
\tabletypesize{\scriptsize} \tablecolumns{9} \tablewidth{0pc}
\tablecaption{Parameters of the continuum sources} \tablehead{
\colhead{MM} &  \colhead{RA Offset\tablenotemark{a}} &  \colhead{DEC Offset\tablenotemark{a}}   & \colhead{ S$_{\nu}$\tablenotemark{b} } & \colhead{R\tablenotemark{c}} & \colhead{n\tablenotemark{d}} &
\colhead{M\tablenotemark{d}}  & \colhead{$\sigma_{vir}$}  & \colhead{M$_{J}$}    \\
\colhead{}  & \colhead{($\arcsec$)}& \colhead{($\arcsec$)} &
\colhead{(10$^{-2}$ Jy)} &
\colhead{(AU)} & \colhead{($10^{7}$ cm$^{-3}$)} &
\colhead{(M$_{\sun}$)}  & \colhead{(km~s$^{-1}$)} & \colhead{(M$_{\sun}$)}
  } \startdata
\multicolumn{9}{c}{SCUBA+Herschel clump} \\
      & ---  & --- & ---  & 36000 &1.6$_{-0.1}^{+0.1}$E-2& 210(12)    & 1.3 & 3.6    \\
\cline{1-9}
\multicolumn{9}{c}{Comp+Ext 1.1 mm core} \\
        & 0.05(0.05) & -0.58(0.06)    & 206.9(9.7) &    4086(196)    & 2.7$_{-0.4}^{+0.4}  $    & 50.9(2.4)   & 1.9 & 0.28   \\
\cline{1-9}
\multicolumn{9}{c}{Ext 1.1 mm condensations} \\
1       &  0.16(0.01) &  0.18(0.02)    & 41.0(1.2) &    1134(35)    & 16.0$_{-1.4}^{+1.6}  $    &6.5(0.2)  & 1.3  &0.21       \\
2       &  0.09(0.01) & -0.95(0.02)    & 39.2(1.2) &     882(40)    & 32.5$_{-4.0}^{+4.9}  $    &6.2(0.2)  & 1.4  &0.15       \\
3       & -0.66(0.03) & -1.95(0.03)    & 14.7(1.0) &     666(60)    & 43.8$_{-10.0}^{+14.3}$    &3.6(0.2)  & 1.3  &0.07       \\
4       &  1.98(0.08) &  0.18(0.05)    &  2.2(0.3) &     666(150)   &  6.6$_{-3.0}^{+7.6}$      &0.5(0.1)  & 0.5  &0.18       \\
5       & -1.67(0.07) & -2.71(0.04)    &  1.4(0.1) & $<$1300        & $>$0.6                    &0.3($<$0.1)  & 0.3  &$<$0.61    \\
6       & -3.37(0.01) & -1.39(0.01)    &  1.8(0.1) & $<$1300        & $>$0.7                    &0.4($<$0.1)  & 0.3  &$<$0.54    \\
7       & -4.41(0.08) & -1.98(0.03)    &  1.7(0.1) & $<$1300        & $>$0.7                    &0.4($<$0.1)  & 0.3  &$<$0.55    \\
8       & -7.24(0.06) & -1.69(0.05)    &  2.2(0.3) &     918(168)    & 2.5$_{-1.0}^{+2.1}$      &0.5(0.1)  & 0.4  &0.29       \\
9       & -6.99(0.07) & -2.41(0.06)    &  2.4(0.3) &    1386(280)    & 0.8$_{-0.3}^{+0.8}$      &0.6(0.1)  & 0.4  &0.51       \\
10      & -8.76(0.03) & -1.24(0.04)    &  1.2(0.1) & $<$1300      & $>$0.5                      &0.3($<$0.1)  & 0.3  &$<$0.66    \\
11      & -8.46(0.02) & -2.22(0.02)    &  1.9(0.1) &    1026(82)     & 1.5$_{-0.3}^{+0.4}$      &0.5($<$0.1)  & 0.4  &0.37       \\
12      &-10.68(0.04) & -5.22(0.02)    &  1.8(0.1) &    1080(174)    & 1.3$_{-0.5}^{+0.9}$      &0.4($<$0.1)  & 0.3  &0.41       \\
13      & -4.22(0.04) & -4.63(0.03)    &  4.2(0.3) &    1602(149)    & 0.9$_{-0.2}^{+0.3}$      &1.0(0.1)  & 0.4  &0.48       \\
14      &  1.05(0.04) & -5.86(0.03)    &  1.9(0.2) &     972(112)    & 1.8$_{-0.5}^{+0.8}$      &0.5(0.1)  & 0.4  &0.34       \\
15      &  3.82(0.01) & -5.57(0.01)    &  0.9(0.1) & $<$1300         & $>$0.4                   &0.2($<$0.1)  & 0.2  &$<$0.76    \\
16      &  3.52(0.02) & -0.85(0.01)    &  1.4(0.1) & $<$1300         & $>$0.6                   &0.3($<$0.1)  & 0.3  &$<$0.61    \\
17      &  1.01(0.04) &  1.99(0.04)    &  1.3(0.1) & $<$1300         & $>$0.5                   &0.3($<$0.1)  & 0.3  &$<$0.63    \\
18      & -3.17(0.08) &  3.86(0.04)    &  1.7(0.2) & $<$1300         & $>$0.7                   &0.4($<$0.1)  & 0.3  &$<$0.55    \\
19      & -2.09(0.07) &  4.15(0.07)    &  3.1(0.4) &    1170(304)    & 1.7$_{-0.8}^{+2.5}$      &0.8(0.1)  & 0.4  &0.35       \\
20      & -2.10(0.04) &  5.58(0.11)    &  3.5(0.3) &    1314(300)    & 1.4$_{-0.6}^{+1.6}$      &0.9(0.1)  & 0.4  &0.39       \\
21      & -0.92(0.08) &  5.04(0.03)    &  4.8(0.3) &    1422(253)    & 1.5$_{-0.6}^{+1.2}$      &1.2(0.1)  & 0.5  &0.38       \\
22      & -0.07(0.24) &  5.19(0.17)    &  3.2(0.3) &    1224(620)    & 1.5$_{-1.1}^{+11.0}$     &0.8(0.1)  & 0.4  &0.37
\enddata
\tablenotetext{a}{Offsets with respect to the phase center R.A.(J2000)~=~05$^{\rm h}$30$^{\rm m}$48.02$^{\rm s}$ and DEC.(J2000)~=~$33\arcdeg47\arcmin54.48\arcsec$.}
\tablenotetext{b}{Total flux from 2-D Gaussian fit. The uncertainties due to flux calibration are $\sim$20\%.}
\tablenotetext{c}{For condensations which are not resolved, we take the beam size as the upper limit for their radii. }
\tablenotetext{d}{We assume a dust temperature of 45 K and a $\beta$ of 1 for MM-1 and MM-2 in calculating masses. For other condensations and the whole clump, we assume a dust temperature of 30 K and a $\beta$ of 1.3. The uncertainties of masses and volume densities are only from flux measurements in 2-D Gaussian fits. The uncertainties of masses and volume densities due to flux calibration are $\sim$20\%, which are not considered.  }
\end{deluxetable*}

The upper panel of Figure 2 shows the 450 $\micron$ continuum emission in pink contours overlayed on the Moment 1 map of the main hyperfine component of NH$_{3}$ (1,1). The emission peak of the 450 $\micron$ continuum emission coincides with the NH$_{3}$ emission peak very well. The NH$_{3}$ emission shows very filamentary and clumpy structures with clear velocity gradients along the filaments. We will discuss the NH$_{3}$ filaments in section 4.4. The black contours in the lower panel of Figure 2 represent the 1.1 mm continuum obtained from combining the compact and extended SMA array data. Three elongated structures (``F$_{N}$", ``F$_{W}$", ``F$_{SE}$") were found in the combined SMA 1.1 mm continuum emission. In section 4.4, we will show that these elongated structures are roughly coincided with the NH$_{3}$ filaments.  With a dust temperature of 30 K and a $\beta$ of 1.3, the total mass derived from the combined SMA 1.1 mm continuum emission is $\sim$51 M$_{\sun}$, which are consistent with the value (50 M$_{\sun}$) derived in \cite{zhang07}. The mean volume density revealed by the 1.1 mm continuum is about 2.7$\times10^{7}$ cm$^{-3}$, which is about two orders of magnitude larger than the mean clump density derived from SCUBA and Herschel continuum emission. The total mass revealed by 1.1 mm continuum is only about one fifth of the clump mass derived from SCUBA and Herschel continuum emission, indicating that about two thirds of the clump gas is distributed in a more extended reservoir.

The 1.1 mm continuum from the SMA extended array data shown as pink contours in the lower panel of Figure 2 reveals a cluster of condensations. The AFGL 5142 clump is highly fragmented along the elongated structures (``F$_{N}$", ``F$_{W}$", ``F$_{SE}$"). As shown in Figure 3, we identified 22 condensations above 4 $\sigma$. The condensations MM 1-5 have been detected in 1.3 mm continuum emission in \cite{zhang07}. Following \cite{zhang07}, we use a dust temperature of 45 K and a $\beta$ of 1 for MM-1 and MM-2 to calculate their masses assuming that the 1.1 mm continuum emission is optically thin. For the other condensations, we use a dust temperature of 30 K and a $\beta$ of 1.3. Their masses and volume densities are listed in Table 1. The masses for MM-1 and MM-2 are about two times larger than those derived in \cite{zhang07} because that \cite{zhang07} only used the peak fluxes to estimate the masses while we used their total fluxes. We obtained the total fluxes from 2-D Gaussian fits. We fitted three 2-D Gaussian functions to separate MM-1, MM-2 and MM-3. The condensations except for MM 1-3 have masses of $\leq$1 M$_{\sun}$. Condensations MM 1-3 have volume densities larger than 1$\times10^{8}$ cm$^{-3}$. The volume densities of other condensations range from 0.5$\times10^{7}$ to 6.6$\times10^{7}$ cm$^{-3}$.

\subsection{Line emission}

\subsubsection{Molecular Lines from KVN observations}

\begin{figure}
\centering
\includegraphics[angle=0,scale=0.4]{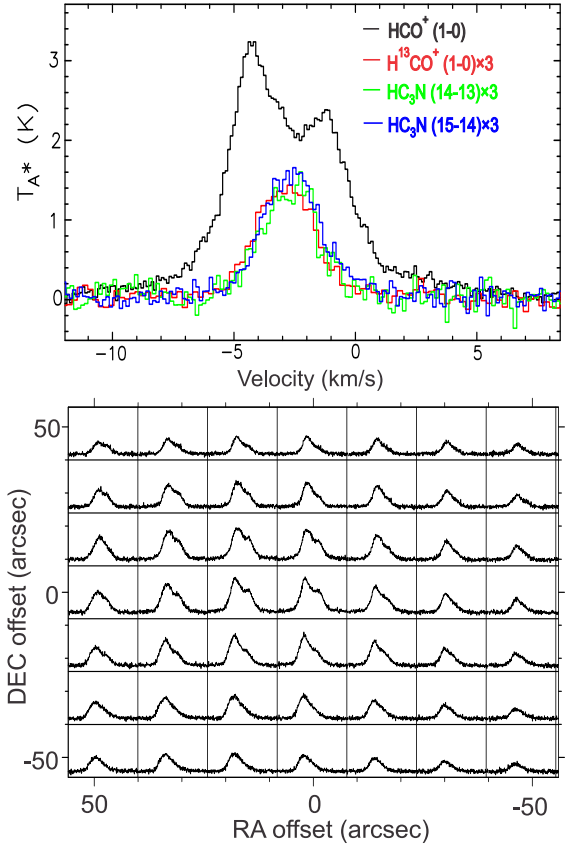}
\caption{ Upper panel: Spectra of HCO$^{+}$ (1-0), H$^{13}$CO$^{+}$ (1-0), HC$_{3}$N (14-13) and (15-14) observed by the KVN telescope at the central position of AFGL 5142.  Lower panel: Grid map of HCO$^{+}$ (1-0) }
\end{figure}

Figure 4 shows the spectra of AFGL 5142 obtained with the KVN 21m telescope. The upper panel presents the spectra at the central position. Optically thin lines ( H$^{13}$CO$^{+}$ (1-0), HC$_{3}$N (14-13) and (15-14) ) have similar single-peaked line profiles. The peak velocities of H$^{13}$CO$^{+}$ (1-0), HC$_{3}$N (14-13) and (15-14) lines from Gaussian fits are -2.89(0.03), -2.69(0.05)and -2.61(0.03) km~s$^{-1}$. We use the mean velocity (-2.7 km~s$^{-1}$) of these three lines as the systemic velocity of AFGL 5142 clump. The line widths of H$^{13}$CO$^{+}$ (1-0), HC$_{3}$N (14-13) and (15-14) lines are 2.96(0.07), 2.87(0.11) and 2.97(0.07) km~s$^{-1}$. The mean line width is $\sim$2.9 km~s$^{-1}$, corresponding to a velocity dispersion of $\sim$1.2 km~s$^{-1}$.

When compared with optically thin lines, HCO$^{+}$ (1-0) line shows a blue asymmetric double-peak line profile (``blue profile") in which the blue emission peak has a higher intensity than the red one. Such a ``blue profile" is a signature for collapse \citep{zhou93}. The lower panel shows the grid map of HCO$^{+}$ (1-0). Across the whole mapping area, all the spectra of HCO$^{+}$ (1-0) show either ``blue profile" or blue asymmetrically skewed profile, indicating that the whole cloud might be in global collapse. Such global collapse has been found in several other high-mass star forming regions \citep{pere12,liu13a,liu13b,qin15}. However, the HCO$^{+}$ (1-0) line emission in previous OVRO interferometric observations does not show a similar ``blue profile" \citep{hunt99}. This may be because that the OVRO interferometric observations filtered out the large scale extended emission and present only the compact structures inside the clump. The single dish KVN observations detect more extended emission from the infalling envelope.

We can derive the infall velocity from KVN HCO$^{+}$ (1-0) line at the central position using the model of \citet{myers96}, in which the infall velocity is given by
\begin{equation}
V_{in}\approx\frac{\sigma^{2}}{v_{red}-v_{blue}}\textrm{ln}\left(\frac{1+eT_{BD}/T_{Dip}}{1+eT_{RD}/T_{Dip}}\right)
\end{equation}
where $T_{Dip}$ is the brightness temperature of the dip (assuming it
is optically thick), and $T_{BD}$ and $T_{RD}$ are the height of the
blue and red peaks above the dip, respectively. The velocity
dispersion $\sigma\sim$1.2 km~s$^{-1}$ is obtained from the
optically thin lines ( H$^{13}$CO$^{+}$ (1-0), HC$_{3}$N (14-13) and (15-14) ). The
infall velocity inferred from HCN (1-0) is $\sim$0.26 km~s$^{-1}$. Assuming spherical symmetry and that the infall radius ($r_{in}$) is the same as the clump radius (R), the mass infall rate ($\dot{M}_{in}$) can be derived as:
\begin{equation}
\begin{split}
\dot{M}_{in}=4\pi r_{in}^2\mu m_{H}nV_{in}=\frac{3M}{R}V_{in}\\ \approx6.3\times10^{-4}(\frac{M}{100~M_{\sun}})(\frac{R}{10^{5}~AU})^{-1}(\frac{V_{in}}{1~km~s^{-1}})M_{\sun}yr^{-1}
\end{split}
\end{equation}
Taking R$\sim36000$ AU and M=210 M$_{\sun}$ as derived from SCUBA and Herschel continuum data, we get an mass infall rate of $\sim$9.6$\times10^{-4}$ M$_{\sun}$yr$^{-1}$, which is about 1.5 times larger than the infall rate ($\sim$6$\times10^{-4}$ M$_{\sun}$yr$^{-1}$) directly measured from CH$_{3}$OH maser emission at radius of 300 AU from the protostar \citep{God11}.

\subsubsection{Spectra and channel maps of molecular lines in SMA observations}

\begin{figure}
\centering
\includegraphics[angle=0,scale=0.45]{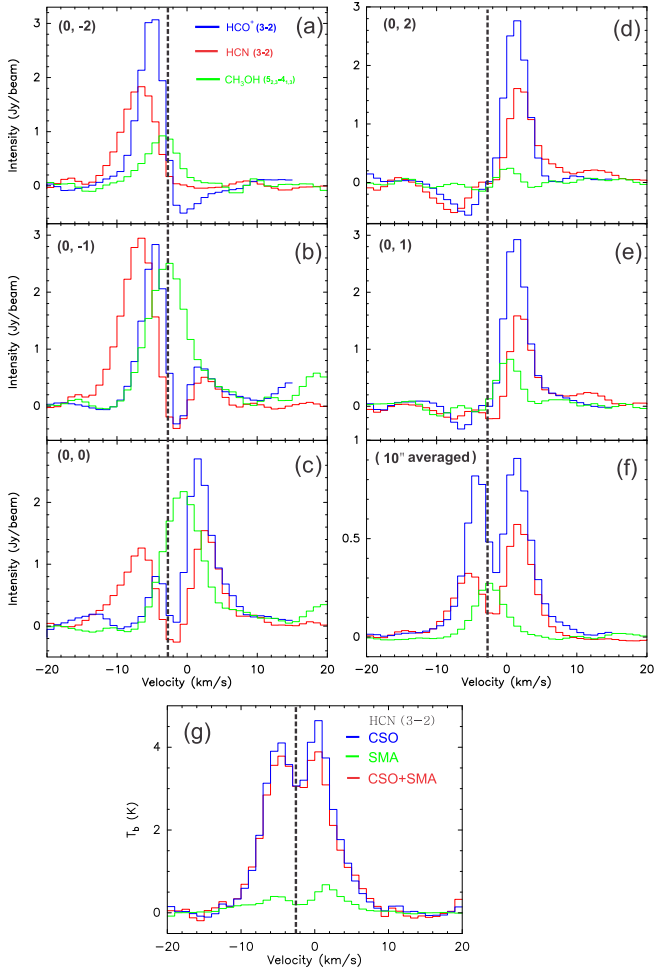}
\caption{SMA spectra at different positions are shown in panels (a) to (e). Panel (f) shows 10$\arcsec$ averaged spectra from the SMA observations. The spectra shown in panels (a) to (f) are: HCO$^{+}$ (3-2) in blue, HCN (3-2) in red and CH$_{3}$OH ($5_{2,3}-4_{1,3}$) in green. In panel (g), we compare HCN (3-2) spectra at the center from SMA and CSO observations. For comparison, all the spectra in panel (g) are convolved to the CSO beam size. HCN (3-2) from the CSO observations only is shown in blue, HCN (3-2) from combined CSO and SMA data in red and HCN (3-2) from the SMA observations only in green. The vertical dashed lines in all panels represent the systemic velocity (-2.7 km~s$^{-1}$) of the clump derived from KVN single-dish observations.    }
\end{figure}

\begin{figure}
\centering
\includegraphics[angle=-90,scale=0.4]{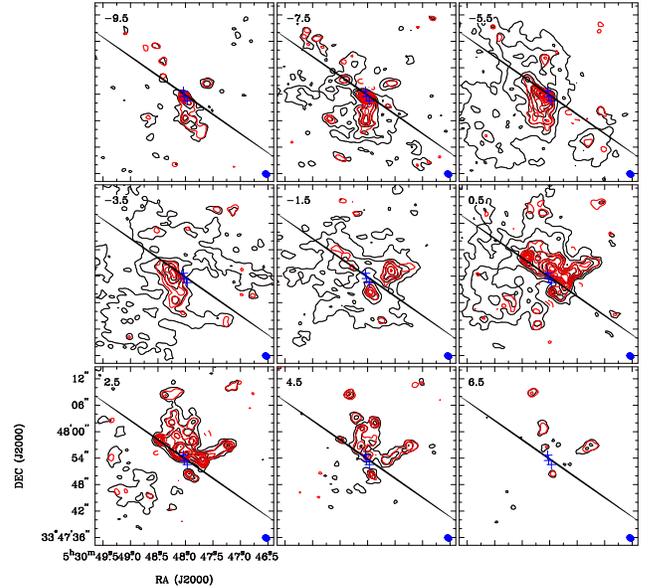}
\caption{Channel maps of HCN (3-2) emission. HCN (3-2) emission from combined CSO and SMA data is plotted in black contours, HCN (3-2) emission from the SMA data only is plotted in red contours. The contour levels are from 0.3 Jy~beam$^{-1}$ in steps of 0.3 Jy~beam$^{-1}$. The three brightest millimeter continuum peaks are marked with blue crosses. The black long line indicates the middle plane of the ``EWBO". The corresponding velocity of
the emission is labeled at the top right corner of each panel.    }
\end{figure}

\begin{figure}
\centering
\includegraphics[angle=-90,scale=0.4]{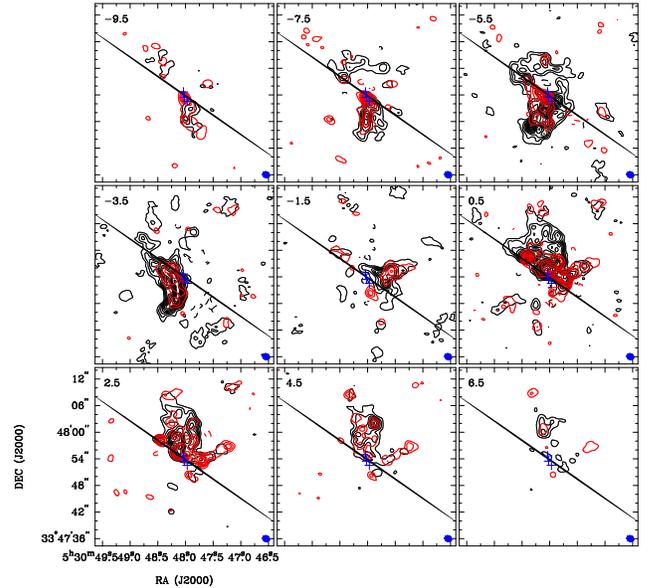}
\caption{Channel maps of SMA HCN (3-2) and HCO$^{+}$ (3-2) emission. The HCN (3-2) and HCO$^{+}$ (3-2) images from the SMA observations are plotted in red and black contours, respectively. The contour levels, crosses and the black long line are the same as in Figure 4. }
\end{figure}

Figure 5 presents the spectra of CH$_{3}$OH ($5_{2,3}-4_{1,3}$), HCN (3-2) and HCO$^{+}$ (3-2) lines. Panels (a) to (e) show the spectra at five positions from the SMA observations. Panel (f) shows the spectra at the center which are averaged within 10$\arcsec$ box area. The CH$_{3}$OH ($5_{2,3}-4_{1,3}$) line spectra at panels (c) and (f) are single peaked.The 10$\arcsec$ averaged CH$_{3}$OH ($5_{2,3}-4_{1,3}$) line with a peak velocity of $\sim$-2 km~s$^{-1}$ is redshifted with respect to the systemic velocity (-2.7 km~s$^{-1}$) of the clump measured from single dish observations, indicating that the core region may have a different systemic velocity than the whole clump. Therefore, we take -2 km~s$^{-1}$ as the systemic velocity of the core region. The HCN (3-2) and HCO$^{+}$ (3-2) line spectra in panels (c) and (f) show double-peak profile with the red emission peak having a slightly higher intensity than the blue one, a ``red profile". The spectra of HCN (3-2) and HCO$^{+}$ (3-2) lines at other positions (panels a, b, d and e) are either totally blueshifted or redshifted with respect to the systemic velocity. In panel (g), we compare HCN (3-2) spectra from SMA, CSO and SMA+CSO observations. The combined CSO and SMA data recovers most of HCN (3-2) emission. Both the CSO and SMA+CSO spectra of HCN (3-2) show a ``red profile", in contrast to the ``blue profile" of the KVN HCO$^{+}$ (1-0) spectrum. The disagreement in line profiles for different lines may be because they trace very different regions of the clump. the HCO$^{+}$ (1-0) line has a critical density about two orders of magnitude lower than that of the HCN (3-2) line \citep{shir15}. Therefore, the HCO$^{+}$ (1-0) line may trace a cold, low density infalling envelope, while the HCN (3-2) line emphasizes a much denser and warmer region.

In Figure 6, we show the channel maps of HCN (3-2) line. The SMA+CSO HCN (3-2) emission reveals more extended structure especially in velocity channels from -7.5 to 2.5 km~s$^{-1}$. However, in high velocity channels (-9.5, 4.5 and 6.5 km~s$^{-1}$), the HCN (3-2) emission in the SMA only data traces very similar structure as in the SMA+CSO combined data, indicating that SMA detected most emission at high velocity. In Figure 7, we compare the channel maps of HCN (3-2) line with those of HCO$^{+}$ (3-2) line. In general, both lines reveal very similar structures. We plot a black long line in each channel. The redshifted and blueshifted emission of both HCN (3-2) and HCO$^{+}$ (3-2) lines can be roughly separated by this line. The emission from -9.5 to -3.5 km~s$^{-1}$ is distributed to the southwest, while the emission from 0.5 to 6.5 km~s$^{-1}$ is mostly to the northwest. Such a bipolar structure could be caused by expansion, rotation or cloud-cloud collision. We can rule out the cloud-cloud collision scenario because there are no hints of two velocity components in optically thin lines. If a cloud-cloud collision were responsible, CH$_{3}$OH ($5_{2,3}-4_{1,3}$) should also show a double peaked profile in the averaged spectrum, which is not seen in panel (f) of Figure 5. Especially, the optically thin lines ( H$^{13}$CO$^{+}$ (1-0), HC$_{3}$N (14-13) and (15-14) ) in single-dish observations are also single-peaked. In section 3.4, we will rule out the rotation scenario. We will argue that the bipolar structure seen in channel maps is most likely due to expansion driven by outflows.

\subsection{High velocity jet-like molecular outflows}

\begin{figure}
\centering
\includegraphics[angle=0,scale=0.45]{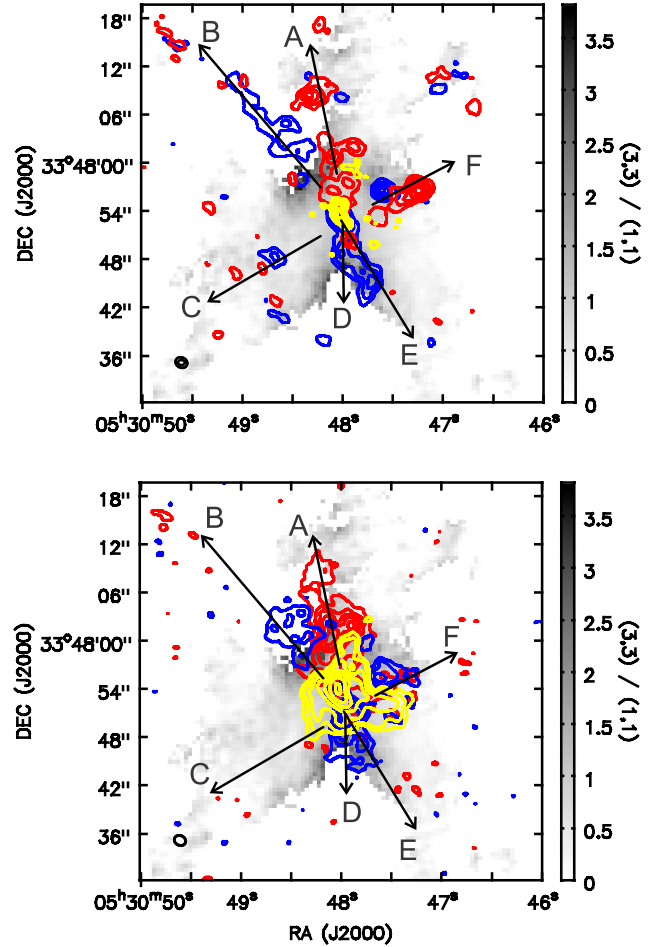}
\caption{Upper panel: HCN (3-2) outflows. Lower panel: HCO$^{+}$ (3-2) outflows. The gray scale images in both panels show the ratio of the NH$_{3}$ (J,K)=(3,3) to (J,K)=(1,1). The 1.1 mm continuum emission from the SMA extended array data is shown in yellow contours in the upper panel. The 1.1 mm continuum emission from the SMA compact and extended array data is shown in yellow contours in the lower panel. The contour levels for 1.1 mm continuum are 5, 10, 20, 40, 80, 160, and 320 $\sigma$. The red and blue contours in upper panel are the integrated intensity of HCN (3-2) emission in the velocity range [4,14] and [-18,-8] km~s$^{-1}$, respectively. The red and blue contours in the lower panel are the integrated intensity of HCO$^{+}$ (3-2) emission in the velocity range [4,14] and [-18,-8] km~s$^{-1}$, respectively. The red and blue contour levels in both panels are from 1 Jy~beam$^{-1}$~km~s$^{-1}$ in steps of 1 Jy~beam$^{-1}$~km~s$^{-1}$. The directions of the high-velocity molecular outflows are indicated by arrows.  }
\end{figure}

As shown in panel (f) of Figure 5, CH$_{3}$OH ($5_{2,3}-4_{1,3}$) emission is mostly from -8 to 4 km~s$^{-1}$. One can easily identify line wings in HCO$^{+}$ (3-2) and HCN (3-2) lines beyond -8 and 4 km~s$^{-1}$ in the spectra. The high velocity emission is also clearly seen in velocity channels (-9.5, 4.5 and 6.5 km~s$^{-1}$) in Figure 6 and 7. We define outflow emission in velocity range [4, 14] and [-18, -8] km~s$^{-1}$ for redshifted and blueshifted outflow components, respectively. The integrated intensity maps of high velocity emission of HCO$^{+}$ (3-2) and HCN (3-2) lines are shown in Figure 8. The orientations and spatial distributions of these outflows are roughly consistent with that of the CO outflows as shown with arrows \citep{zhang07}. We named the outflows ``A" to ``F". The outflows revealed by the HCO$^{+}$ (3-2) and HCN (3-2) lines are very clumpy. Outflow ``A" is dominated by redshifted emission. Outflows ``B, C, D, E" are dominated by blueshifted emission. Outflow ``F" shows both redshifted and blueshifted emission. Outflow ``C" shows weak emission in HCN (3-2) but no emission in HCO$^{+}$ (3-2). In general, HCN (3-2) traces more compact and narrower jet-like outflows when compared with HCO$^{+}$ (3-2), which traces more extended and wider outflow structures. Especially for outflow ``B", HCN (3-2) reveals a long and narrow jet-like outflow.  HCN (3-2) has a critical density more than five times larger than that of HCO$^{+}$ (3-2), indicating that HCN (3-2) could more easily trace dense gas close to the outflow jets than HCO$^{+}$ (3-2) does.

The gray-scale images in Figure 8 show the ratio of the NH$_{3}$ (J,K)=(3,3) to (J,K)=(1,1). Along the outflows ``A, B, D, F", NH$_{3}$ emission shows cavity structure, which is suggestive of outflow/evelope interaction \citep{zhang02}. The yellow contours in Figure 8 represent the 1.1 mm continuum emission. The three elongated structures especially ``F$_{w}$" of continuum emission are mainly located between the outflows, probably indicating outflow/core interaction. The three densest condensations MM 1-3 are located at the center of outflows. However, with present data we could not tell which condensations are the driven sources for the outflows.

\subsection{The extremely wide-angle bipolar outflow (EWBO)}

\begin{figure*}
\centering
\includegraphics[angle=-90,scale=0.5]{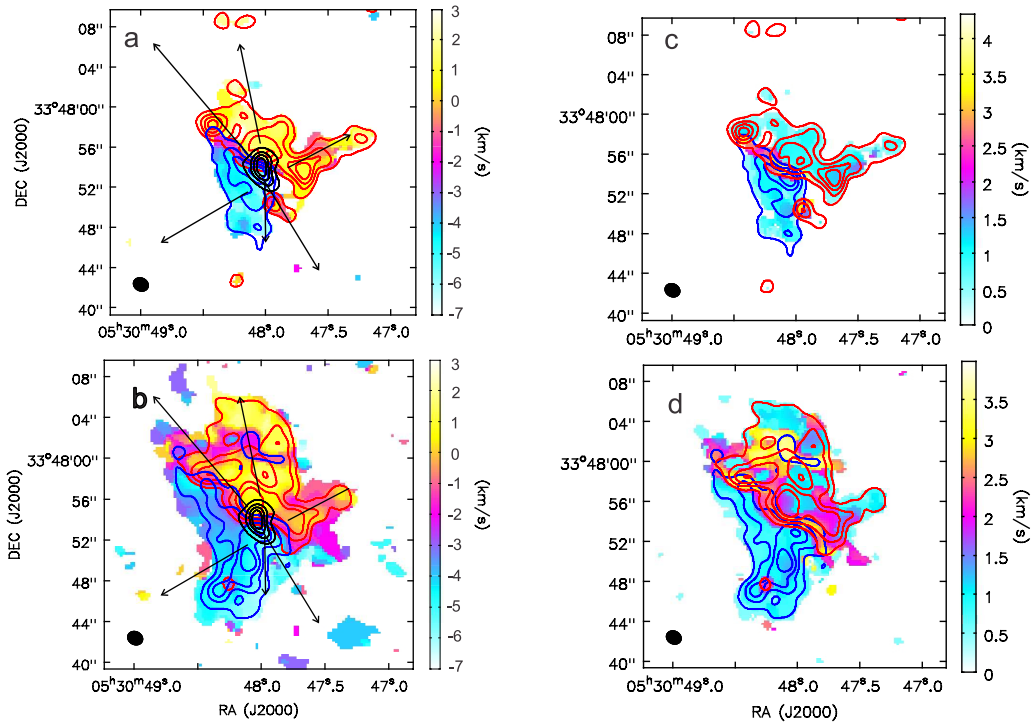}
\caption{The Moment 1 maps of HCN (3-2) and  HCO$^{+}$ (3-2) emission from the SMA observations in the velocity range [-7,3] km~s$^{-1}$ are shown in color scale in panels (a) and (b), respectively. The Moment 2 maps of HCN (3-2) and  HCO$^{+}$ (3-2) emission in the velocity range [-7,3] km~s$^{-1}$ are shown in color scale in panels (c) and (d), respectively. The red and blue contours in panels (a) and (c) are integrated intensity of HCN (3-2) emission in the velocity range [-1,3] and [-7,-3] km~s$^{-1}$, respectively. The red and blue contours in panels (b) and (d) are integrated intensity of HCO$^{+}$ (3-2) emission in the velocity range [-1,3] and [-7,-3] km~s$^{-1}$, respectively. The contour levels in panels (a), (b), (c) and (d) are from 20\% to 80\% in steps of 20\% of the peak values. The peak values of the red and blue contours of HCN (3-2) are 7.86 and 8.57 Jy~beam$^{-1}$~km~s$^{-1}$, respectively. While the peak values of the red and blue contours of HCO$^{+}$ (3-2) are 12.85 and 14.68 Jy~beam$^{-1}$~km~s$^{-1}$, respectively. The black contours in panels (a) and (b) show the 1.1 mm continuum emission from the combined SMA compact and extended array data. The contour levels are from 20\% to 80\% in steps of 20\% of the peak value (0.526 Jy~beam$^{-1}$). The directions of high-velocity molecular outflows are indicated by arrows. }
\end{figure*}

From low velocity channels between -5.5 and  2.5 km~s$^{-1}$ in Figure 6 and 7, we find that the HCN (3-2) and HCO$^{+}$ (3-2) emission are roughly blueshifted in the south-east region and redshifted in the north-west region. This structure is more clearly seen in the Moment 1 maps of HCN (3-2) and HCO$^{+}$ (3-2) lines in panels (a) and (b) of Figure 9, in which the redshifted and blueshifted emission are clearly separated. We define the low-velocity emission in the velocity range of (-7,3) km~s$^{-1}$ to avoid contamination from high velocity emission.  The integrated intensities of the blueshifted (from -7 to -3 km~s$^{-1}$) and redshifted (from -1 to 3 km~s$^{-1}$) low-velocity emission were shown as contours in Figure 9. As the same in the channel maps, the blueshifted integrated intensity contours are mainly located to south-east and the redshifted contours to north-west. We argue that this low-velocity ``blob like" bipolar structure is not caused by rotation because:

(a). We estimate the characteristic flow velocities by averaging the pixel values in Moment 1 maps and present those values in column 3 of Table 2. The mean flow velocity is $\sim$3 km~s$^{-1}$, which is a lower limit because we do not take into account the inclination. The mean radius of the ``blob like" structure is $\sim$0.08 pc ($\sim$16500 AU). If it is caused by rotation, the dynamical mass of the central object should be $M=\frac{Rv^{2}}{G}\geq170 M_{\sun}$, which is much larger than the total mass ($\sim$51 M$_{\sun}$) measured from 1.1 mm continuum, indicating that this structure could not be supported by rotation.

(b). From the channel maps in Figure 6 and 7, one can see that the low-velocity ``blob like" structure is greatly shaped by the high-velocity outflows. It becomes more apparent in the -7.5, 2.5 and 4.5 km~s$^{-1}$ channels, in which the low-velocity ``blob like" structure is roughly elongated in the orientation of the high-velocity outflows. Therefore, we suggest that low-velocity ``blob like" structure is more like envelope material entrained by jet-like high-velocity outflows. The base of the low-velocity ``blob like" outflow has an opening angle of nearly $\sim$180$\arcdeg$. This extremely wide-angle bipolar outflow is denoted as ``EWBO" hereafter.

Panels (c) and (d) of Figure 9 present the Moment 2 maps of the HCO$^{+}$ (3-2) and HCN (3-2) lines. The Moment 2 maps show very smooth and coherent structure of the ``EWBO". The velocity dispersion is artificially large at the interface of the red and blue low-velocity outflow due to the overlap of the redshifted and blueshifted lobes. The average velocity dispersion of each lobe is presented in column 4 of Table 2. The mean velocity dispersion of the ``EWBO" is $\sim$0.9 km~s$^{-1}$.

\section{Discussion}

\subsection{Grain growth toward the clump center?}

\begin{figure*}
\centering
\includegraphics[angle=90,scale=0.6]{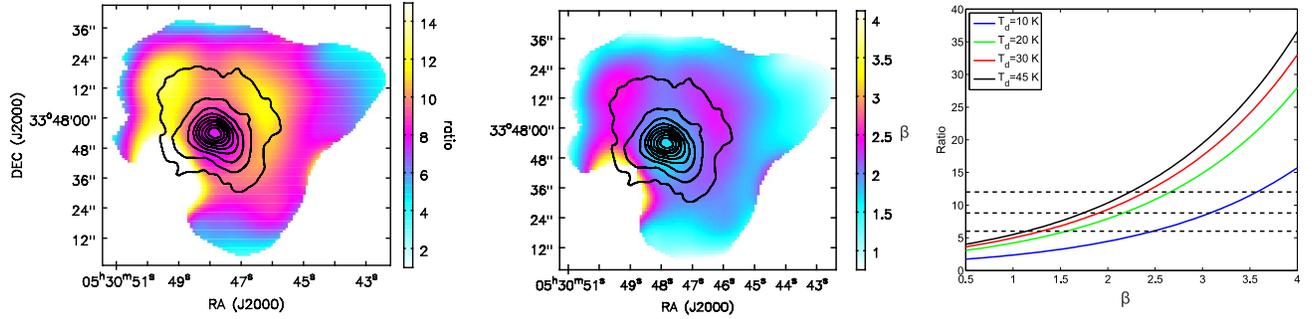}
\caption{Left panel: The SCUBA 450 $\micron$/850$\micron$ flux ratio is shown in color image. The flux intensity of 450 $\micron$ continuum is shown in contours. The contour levels are from 10\% to 90\% in steps of 10\% of the peak value 45.5 Jy/beam. Middle panel: The emissivity spectral index ($\beta$) map is shown in color scale. The contours are the same as in the left panel. Right panel: The 450 $\micron$/850$\micron$ flux ratio as function of emissivity spectral index for different dust temperatures is shown as solid lines. From top to bottom, the dashed lines represent 450 $\micron$/850 $\micron$ flux ratio equal to 12, 8.8 and 6, respectively.  }
\end{figure*}

In the left panel of Figure 10, we present the 450 $\micron$ continuum emission in contours overlayed on the 450 $\micron$/850 $\micron$ flux ratio color image. We smoothed the 450 $\micron$ and 850 $\micron$ continuum data with a Gaussian function to the same 17$\arcsec$.4 resolution to construct the flux ratio map. Assuming that the dust emission is optically thin, the 450 $\micron$/850$\micron$ flux ratio can be expressed as a function of the dust temperature (T$_{d}$) and emissivity spectral index ($\beta$), using the following formulae \citep{zhu09}:
\begin{equation}
\frac{S_{450}}{S_{850}}=1.88^{\beta+3}(\frac{e^{16.8/T_{d}}-1}{e^{31.8/T_{d}}-1})
\end{equation}

We consider the possibility that the 450 $\micron$ emission may have high opacity toward the more dense central
region. Assuming a dust temperate of 30 K, we derived an optical depth of 0.08 for 450 $\micron$ emission using its peak flux of 45.5 Jy/beam in equation (1). The optical depth will be even smaller if we take a higher dust temperature. Therefore, the 450 $\micron$ emission is apparently optically thin and the 450 $\micron$/850$\micron$ flux ratio should not be affected by optical depth.

Interestingly, the 450 $\micron$/850 $\micron$ flux ratio increases from the center to a radius of $\sim$22$\arcsec$ and then decreases to the clump edge. The 450 $\micron$/850$\micron$ flux ratio is greatly enhanced at a radius of $\sim$22$\arcsec$ (envelope region) with a mean value of $\sim$12. The mean 450 $\micron$/850$\micron$ flux ratio within $\sim$15$\arcsec$ radius (core region) is about 8.8. The 450 $\micron$/850$\micron$ flux ratio reaches a minimum $\leq$6 near the clump edge. Assuming a constant dust temperature of 30 K as derived from the SED fit in section 3.1.1, we derived the emissivity spectral index ($\beta$) map as shown in the middle panel. It is clearly seen that the central region has smaller $\beta$ than that of the regions at the radius of $\sim$22$\arcsec$, indicating that the core region may have larger dust grains than the envelope.

However, the assumption of a constant dust temperature may be not true since the core region close to protostars may have higher temperature than the envelope. In the right panel of Figure 10, we investigate how the 450 $\micron$/850$\micron$ flux ratio changes with T$_{d}$ and $\beta$. For a given $\beta$, the 450 $\micron$/850$\micron$ flux ratio increases with T$_{d}$. Similarly, for a given T$_{d}$, the 450 $\micron$/850$\micron$ flux ratio increases with $\beta$. For a dust temperature of 30 K, the corresponding $\beta$ for the 450 $\micron$/850$\micron$ flux ratios of 8.8 and 12 are 1.9 and 2.4, respectively.  However, since the core region may be warmer than the envelope, a smaller $\beta$ is needed to explain the low 450 $\micron$/850$\micron$ flux ratio in the core region. Assuming a dust temperature of 45 K, $\beta$ in the core region should be $\sim$1.7. Additionally, since the SCUBA continuum emission is sensitive to both the core and the envelope components, $\beta$ in the core region may be even smaller if one removes the envelope component. Indeed, a small $\beta$ of 1 was found toward the central dense cores from interferometric observations \citep{hunt99,zhang07}. Small $\beta$ suggests large dust grains. The observational evidence may indicate grain growth from envelope to central cores, which needs to be tested by further high angular resolution multi-bands observations.

\subsection{Dynamical State of continuum sources}

To examine the gravitational stability of the continuum sources (clump, core and condensations) in Table 1, we calculate the one dimensional
velocity dispersion ($\sigma_{vir}$) for virial equilibrium between gas motions and gravity:
\begin{equation}
\sigma_{vir}=\sqrt{\frac{\alpha MG}{5R}}.
\end{equation}
$\alpha$ is a factor equal to unity for a
uniform density profile and 5/3 for an inverse square profile. Here we took $\alpha$ equal to 5/3. The derived one dimensional
virial velocity dispersions ($\sigma_{vir}$) are listed in the 8th column of Table 1. The clump revealed in SCUBA continuum emission has a virial velocity dispersion consistent with the
velocity dispersion (1.2 km~s$^{-1}$) measured from optically thin lines ( H$^{13}$CO$^{+}$ (1-0), HC$_{3}$N (14-13) and (15-14) ), indicating that the clump is in virial equilibrium. From a Gaussian fit toward the 10$\arcsec$ averaged spectrum of CH$_{3}$OH ($5_{2,3}-4_{1,3}$) line, we derived a velocity dispersion of $\sim$2.2 km~s$^{-1}$, which is consistent with the virial velocity dispersion of the core revealed in SMA Compact+Extended 1.1 mm continuum emission, indicating that the central core within 4000 AU radius is also in virial equilibrium. However, we do not have direct measurements of line velocity dispersions toward most condensations because the sensitivity of the extended array observations is not good enough to detect molecular lines toward condensations except MM-1 and MM-2. Additionally, the present spectral resolution (0.9 km~s$^{-1}$) is not good enough to accurately measure linewidths smaller than 1 km~s$^{-1}$. With a spectral resolution of 0.42 km~s$^{-1}$, \cite{bus11} find that the velocity dispersions of N$_{2}$H$^{+}$ (1-0) in CARMA observations range from 0.3 to 1 km~s$^{-1}$. These values are consistent with the virial velocity dispersions of most condensations, indicating that most condensations may be also gravitationally bound.

The virial analysis presented above does not include magnetic fields. Magnetic fields can provide additional pressure to increase virial masses. However, from the above analysis, the continuum sources (clump, core and condensations) seem to be in virial equilibrium without considering magnetic fields, indicating that magnetic fields may not play an important role. Additionally, the AFGL 5142 clump is highly fragmented. Recent simulations suggest that high magnetic field values can significantly suppress fragmentation \citep{myers13,myers14}. A strong magnetic field can not explain the high fragmentation level in AFGL 5142, as also suggested by \cite{pal13}. Therefore, turbulence seems to play a more important role than magnetic field in mass assembly of AFGL 5142.

What determines the fragmentation process? We calculated the thermal Jeans mass with the formula:

\begin{equation}\label{eq_lacc}
M_{J}=\frac{\pi^{\frac{5}{2}}c_{s}^{3}}{6\sqrt{G^{3}\rho}}=0.0147M_{\sun}(\frac{T}{10 K})^{\frac{3}{2}}(\frac{M}{10 M_{\sun}})^{-\frac{1}{2}}(\frac{R}{1000 AU})^{\frac{3}{2}}.
\end{equation}

We used 45 K for MM-1 and MM-2 and 30 K for other sources. The derived Jeans masses are presented in column 9 of Table 1. The condensations except for the central condensations MM 1-3 have masses comparable to their thermal Jeans masses or slightly larger by a factor smaller than 3, indicating that the fragmentation in lower density environments away from the center could be determined by thermal pressure. However, thermal Jeans fragmentation cannot explain the masses of MM 1-3. If we take account of the turbulent pressure by replacing the sound speed $C_{s}$ with a velocity dispersion of 1.2 km~s$^{-1}$, the Jeans masses for MM-1, MM-2 and MM-3 will be 5.9, 4.1 and 3.5 M$_{\sun}$, respectively, which are consistent with their masses derived from 1.1 mm continuum. Therefore, this suggests that turbulence
dominates over thermal pressure in suppressing further fragmentation in the central region.

\subsection{Properties of the ``EWBO"}

The high-velocity outflows seem to drive the whole envelope within 0.1 pc from the protostars into expansion, which forms the low-velocity ``blob like" ``EWBO". The ``EWBO" is very clumpy and seems to be a collection of multiple low-velocity outflows. The mean flow velocity is $\sim$3 km~s$^{-1}$. Interestingly, the spectra in panels (d) and (e) of Figure 5 show blueshifted absorption around -6 km~s$^{-1}$. The absorption dips are about 3.3 km~s$^{-1}$ blueshifted with respect to the systemic velocity (-2.7 km~s$^{-1}$) of the clump. The absorption dips could be caused by the foreground cold expanding gas with a velocity of $\sim$3.3 km~s$^{-1}$. The expansion velocity measured from absorption dips are consistent with the flow velocity of the ``EWBO" derived from Moment 1 maps.

In Table 2, we present the parameters of the ``EWBO". The radius and dynamical time of the ``EWBO" are $\sim$0.08 pc and $\sim$2.6$\times10^{4}$ yr, respectively. The dynamical time is consistent with the dynamical time of the high-velocity CO outflows ($\sim10^{4}$ yr) \citep{zhang07}. \cite{bus11} obtained an averaged H$^{13}$CO$^{+}$ column density of 4.5$\times10^{12}$ cm$^{-2}$ toward AFGL 5142. Assuming [HCO$^{+}$/H$^{13}$CO$^{+}$]=60 and a HCO$^{+}$ fractional abundance of 6.4$\times10^{-9}$, which is the median value for protostellar clumps in MALT90 survey \citep{hoq13}, we can estimate the mass of ``EWBO" as:
\begin{equation}
M=\mu m_{H}N_{H_{2}}\pi R^{2}
\end{equation}
where $\mu$=2.37, m$_{H}$, and N$_{H_{2}}$ are mean molecular weight, atomic hydrogen mass and molecular hydrogen column density, respectively. The total mass of the ``EWBO" is 16 M$_{\sun}$, which is about two times larger than that ($\sim$8.5 M$_{\sun}$) of the CO jet-like outflows \citep{zhang07}. The total momentum of the ``EWBO" is $\sim$48 M$_{\sun}$~km~s$^{-1}$, which is consistent with the total momentum (52 M$_{\sun}$~km~s$^{-1}$) of high-velocity CO outflows \citep{zhang07}, indicating that the ``EWBO" could be formed due to the momentum feedback of high velocity outflows. The total energy of the ``EWBO" is 1.4$\times10^{45}$ ergs, which is about one fourth of the total energy ($\sim$5.1$\times10^{45}$ ergs) of the high velocity outflows. The mass loss rate of the ``EWBO" is 6.2$\times10^{-4}$ M$_{\sun}$~yr$^{-1}$, which is comparable to the mass infall rate ($\sim$6$\times10^{-4}$ M$_{\sun}$) of the inner ($<400$ AU) envelope directly measured from CH$_{3}$OH maser emission \citep{God11}. The total mass loss rate of high velocity outflows is $\sim$ 1.1$\times10^{-3}$ M$_{\sun}$~yr$^{-1}$ \citep{zhang07}. Therefore, the total mass loss rate due to both the ``EWBO" and high velocity outflows is 1.7$\times10^{-3}$ M$_{\sun}$~yr$^{-1}$, which is larger than the mass infall rate (9.6$\times10^{-4}$ M$_{\sun}$~yr$^{-1}$) of the clump as measured from HCO$^{+}$ (1-0), indicating that the accreted material may be 100\% converted to outflows or that the protostars may stop accreting if we only consider spherical infall.

The ``EWBO" do have considerable mass and significant momentum and energy, which are comparable to the high-velocity outflows. It is hard to understand the physics by which the narrow-angle high velocity outflows entrain material over a much wider range of angles, and produces the very ``blob like" low velocity outflow. One explantation is that the orientation of outflow jets varies with time and the jet-like high-velocity outflows sweep the whole envelope out. Jet precession could be caused by dynamical interactions between members in the protobinary system of AFGL 5142 \citep{zhang07}.  The discovery of such ``EWBOs" do provide new insights into the interaction between envelopes and high-velocity molecular outflows in high-mass star (or cluster) forming regions. Such interaction in the manner of ``EWBOs" is more energetic and takes place at a much larger spatial scale ($\sim$0.2 pc) in high-mass star forming regions than in their low-mass counterparts. Studies of such rare ``EWBOs" could drastically alter our picture of outflow feedback in high-mass star forming regions.

\begin{deluxetable*}{ccccccc}
\tabletypesize{\scriptsize} \tablecolumns{6} \tablewidth{0pc}
\tablecaption{Parameters of the EWBO} \tablehead{
\colhead{Component} & \colhead{Velocity interval} & \colhead{V$_{char}$\tablenotemark{a} } & \colhead{$\sigma_{1D}$ } & \colhead{L$_{eff}$\tablenotemark{a}} &
\colhead{T$_{dyn}$}   \\
\colhead{} & \colhead{(km~s$^{-1}$)} &
\colhead{(km~s$^{-1}$)} &
\colhead{(km~s$^{-1}$)} &
\colhead{(pc)} &
\colhead{($10^{4}$yr)} &
  } \startdata
HCN Red   &[-1,3]        &    3.7 &  0.8 &  0.07  &     1.9       \\
HCN Blue  &[-7,-3]       &    2.7 &  0.8 &  0.06  &     2.2     \\
HCO$^{+}$ Red &[-1,3]    &    3.3 &  1.0 &  0.09  &     2.7   \\
HCO$^{+}$ Blue &[-7,-3]  &    2.2 &  0.9 &  0.08  &     3.6
\enddata
\tablenotetext{a}{Parameters are not corrected for orientation}
\end{deluxetable*}

\subsection{Hierarchical network of filaments in NH$_{3}$ (1,1) emission}

\begin{figure}
\centering
\includegraphics[angle=0,scale=0.45]{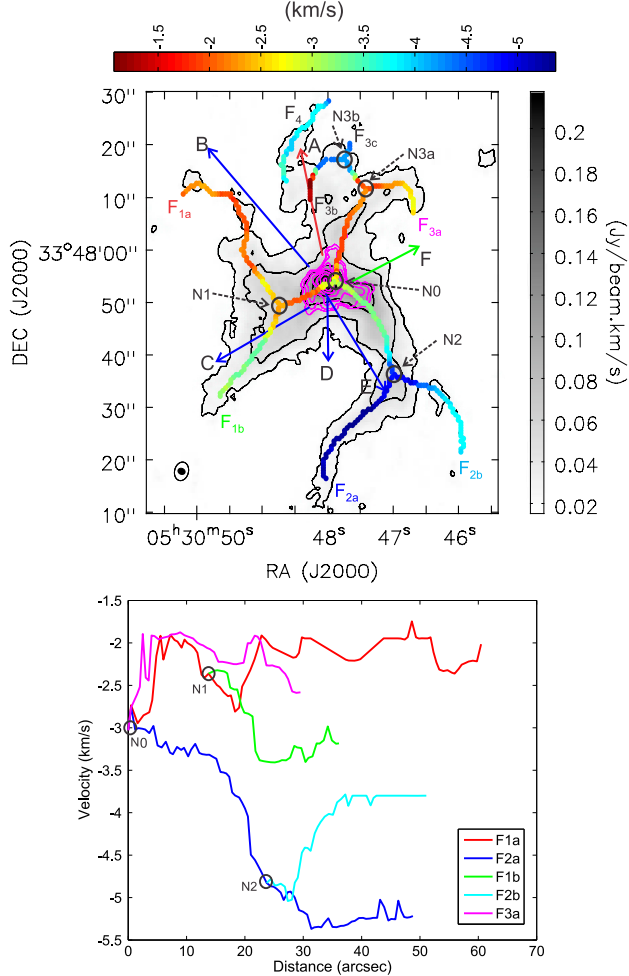}
\caption{Upper panel: The integrated intensity of the main hyperfine component of NH$_{3}$ (1,1) is shown as gray scale image and black contours. The contour levels are 10\%, 20\%, 40\%, 60\% and 80\% of the peak value (0.22 Jy~beam$^{-1}$~km~s$^{-1}$). The 1.1 mm continuum emission from the SMA compact and extended array data are shown in pink contours. The contour levels for 1.1 mm continuum are 5, 10, 20, 40, 80, 160, and 320 $\sigma$. The colored curves show the filaments, color-coded by peak velocities. The directions of high-velocity molecular outflows are indicated by arrows. Lower panel: Position-Velocity diagrams of the main hyperfine component of NH$_{3}$ (1,1) along filaments. The distances were measured from node ``N0" along the filaments.  }
\end{figure}

As mentioned in section 3.1.2, NH$_{3}$ (1,1) emission shows a network of filaments (also see Figure 2). We used the ``FILFINDER" algorithm \citep{koch15} for detecting filamentary structure in NH$_{3}$ (1,1) emission above 3 $\sigma$. When compared to other algorithms, ``FILFINDER" can not only identify the same bright filaments but also reliably extract a population of faint filaments \citep{koch15}. In the upper panel of Figure 11, we plot the skeleton of the filament network, color-coded with the peak velocities of NH$_{3}$ (1,1) emission. We found that the network of filaments is hierarchical with filaments merging at some points (i.e., nodes). In total, we identified eight filaments (``F$_{1a}$" to ``F$_{4}$") and five nodes (``N0" to ``N$_{3b}$"). The names of filaments and nodes are labeled in the upper panel of Figure 11. ``F$_{1a}$" and ``F$_{1b}$" meet at ``N1". ``F$_{2a}$" and ``F$_{2b}$" encounters at ``N2". ``F$_{3b}$" and ``F$_{3c}$" merge at ``N$_{3b}$", and then merge with ``F$_{3a}$" at ``N$_{3a}$". Node ``N0" is close to the continuum emission peak and is connected with three main filaments. Interestingly, we find that the orientations of the three elongated structures in 1.1 mm continuum identified in section 3.1.2 are roughly consistent with the directions of NH$_{3}$ filaments.

One can also see clear velocity gradients along the filaments. In the lower panel of Figure 11, we investigate the velocity drifts along filaments for five main filaments. ``F$_{1a}$" is the longest filament with a length of $\sim$0.52 pc (60$\arcsec$). From ``N0" to 0.07 pc (8$\arcsec$), the velocity of ``F$_{1a}$" drastically changes from $\sim$-3 km~s$^{-1}$ to $\sim$-2 km~s$^{-1}$, leading to a large velocity gradient of $\sim$17 km~s$^{-1}$~pc$^{-1}$. There is another dip in the Position-Velocity diagram of ``F$_{1a}$" at $\sim$0.17 pc ($\sim20\arcsec$) away from node ``N0". From 0.07 pc to 0.17 pc, the velocity gradient of ``F$_{1a}$" is around $\sim$-9 km~s$^{-1}$~pc$^{-1}$. From 0.17 pc to 0.21 pc, the velocity gradient of ``F$_{1a}$" is around $\sim$20 km~s$^{-1}$~pc$^{-1}$. Beyond 0.21 pc away from node ``N0", ``F$_{1a}$" has a roughly constant velocity of $\sim$-2 km~s$^{-1}$ with a variation smaller than $\sim$0.5 km~s$^{-1}$. Filament ``F$_{1b}$" encounters with ``F$_{1a}$" at Node ``N1". From ``N1" to 0.2 pc (23$\arcsec$), ``F$_{1b}$" shows a large velocity gradient of $\sim$-15 km~s$^{-1}$~pc$^{-1}$. Beyond 0.2 pc, ``F$_{1b}$" has a roughly constant velocity of $\sim$-3.3 km~s$^{-1}$. ``F$_{3a}$" has similar velocity gradient ($\sim$17 km~s$^{-1}$~pc$^{-1}$) from node ``N0" to 0.07 pc as ``F$_{1a}$". Filament ``F$_{2a}$" has a length of $\sim$0.44 pc (50$\arcsec$). From Node ``N0" to 0.26 pc (30$\arcsec$), the velocity of ``F$_{2a}$" continuously decreases from -3 to -5.4 km~s$^{-1}$, corresponding to a velocity gradient of $\sim$-10 km~s$^{-1}$~pc$^{-1}$. Beyond 0.26 pc, ``F$_{2a}$" has a roughly constant velocity of $\sim$-5.3 km~s$^{-1}$. In contrast to ``F$_{2a}$", the velocity of ``F$_{2b}$" increases from node ``N2" to 0.31 pc ($\sim36\arcsec$) with a velocity gradient of $\sim$15 km~s$^{-1}$~pc$^{-1}$. The velocity of ``F$_{2b}$" has a roughly constant velocity of $\sim$-3.8 km~s$^{-1}$ beyond 0.31 pc.

Velocity gradients along filaments have been detected in other cluster forming regions, including SDC335 cluster \citep{pere12} and Serpens South cluster \citep{kirk13}. Such velocity gradients of the order of $\sim$1 km~s$^{-1}$~pc$^{-1}$ were interpreted as gas inflow along filaments \citep{pere12,kirk13}. When compared with their results, the velocity gradients along NH$_{3}$ filaments (like ``F$_{1b}$", ``F$_{2a}$" and ``F$_{2b}$") in AFGL 5142 are about one orders of magnitude larger. ``F$_{1a}$" and ``F$_{3a}$" in general show constant velocities along most parts of the filaments, suggesting small velocity gradients. However, this may be caused by projection effects if ``F$_{1a}$" and ``F$_{3a}$" lie parallel to the plane of the sky. Are the velocity gradients caused by rotation? If rotation, Filament ``F$_{1a}$" has a rotation velocity of 2.4 km~s$^{-1}$ at 0.26 pc radius. Then the dynamical mass of the central cluster to bind the system should be $\sim350 M_{\sun}$, which is much larger than the total cluster mass ($\sim51 M_{\sun}$) derived from 1.1 mm continuum. Thus rotation cannot explain the large velocity gradients. However, we cannot ignore the affect of outflow shocks, which may also induce large velocity gradients along the filaments. For example, the end of filament ``F$_{3b}$" has large redshifted velocities, hinting for compression from the red outflow ``A". The large velocity gradient around node ``N2" may be the consequence of outflow-envelope interaction. However, the outflows cannot explain the overall velocity patterns in the filaments. For example, filament ``F$_{1a}$" is close to the blue outflow ``B" but has redshifted velocities. In contrast, filament ``F$_{4}$" is close to the red outflow ``A" but has blueshifted velocities. Additionally, the outflows are generally located between filaments. Therefore, we argue that the velocity gradients are mainly caused by inflow, though rotation and outflow may be making some contributions.

If the velocity gradients are mainly caused by inflows, we can estimate the accretion rate ($\dot{M}_{\parallel}$) along filaments following \cite{kirk13}:
\begin{equation}
\dot{M}_{\parallel} = \frac{\nabla V_{\parallel,obs} M}{tan(\theta)}
\end{equation}
where $\nabla V_{\parallel,obs}$, M and $\theta$ are the velocity gradient, mass and inclination angle to the plane of the sky of filaments, respectively. It is hard to estimate the total mass in filaments due to the uncertainties of NH$_{3}$ abundance. \cite{zhang05} obtained a total mass of 660$\times(10^{-7}/\chi)$ traced by NH$_{3}$, where $\chi$ is the NH$_{3}$ abundance. The maximum NH$_{3}$ abundance for AFGL 5142 reached in the chemical model of \cite{bus11} is $5.2\times10^{-8}$. Assuming $\chi=5.2\times10^{-8}$, the total mass is $\sim$1270 M$_{\sun}$, much larger than the clump mass ($\sim$210 M$_{\sun}$) estimated from SCUBA and Herschel continuum emission. We will use the total clump mass ($\sim$210 M$_{\sun}$) as a lower limit for the total mass trapped in all filaments. From previous analysis, we learned that $\nabla V_{\parallel,obs}$ is of the order of 10 km~s$^{-1}$~pc$^{-1}$.  Assuming $\nabla V_{\parallel,obs}$=10 km~s$^{-1}$~pc$^{-1}$, M=210 M$_{\sun}$ and $\theta=45\arcdeg$, we derive a total accretion rate of $\sim2.1\times10^{-3}$ M$_{\sun}$~yr$^{-1}$ along filaments. The $\dot{M}_{\parallel}$ is about two times the mass infall rate ($\dot{M}_{in}$=$\sim$9.6$\times10^{-4}$ M$_{\sun}$~yr$^{-1}$) along the line of sight calculated from HCO$^{+}$ (1-0). Interestingly, $\dot{M}_{\parallel}$ is comparable to the total mass loss rate ($\sim1.7\times10^{-3}$ M$_{\sun}$yr$^{-1}$) of jet-like outflows and the ``EWBO". The sum of $\dot{M}_{\parallel}$ and $\dot{M}_{in}$ is $\sim$3.1$\times10^{-3}$ M$_{\sun}$~yr$^{-1}$, which overcomes the total mass loss rate, indicating that the central cluster is probably still gaining mass.

\section{Summary}

\begin{figure}
\centering
\includegraphics[angle=-90,scale=0.35]{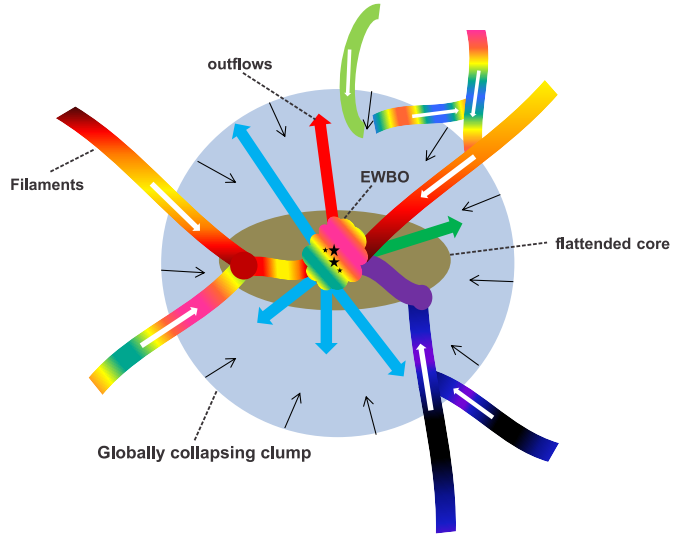}
\caption{Schematic view of AFGL 5142. The whole clump is in global collapse. The central cluster (stars) is embedded in a flattened core (filled ellipse) \citep{zhang05}. The central cluster (``stars") is still accreting gas from the clump/cloud through filaments (``color bands"). The cluster eject masses through both high velocity jet-like outflows (``color arrows") and low velocity EWBO (``color cloud-like object").   }
\end{figure}

In Figure 12, we summarize the observations in a schematic model. The main results of this study are summarized as follows:

1. The AFGL 5142 clump has a mass of $\sim$210 M$_{\sun}$ and a radius of $\sim$36000 AU as measured from JCMT/SCUBA and Herschel continuum emission. The clump is in global collapse with a infall velocity of $\sim$0.26 km~s$^{-1}$ and a mass infall rate of $\sim$9.6$\times10^{-4}$ M$_{\sun}$~yr$^{-1}$ along the line of sight.

2. The central cluster resides in a flattened core as revealed in NH$_{3}$ emission \citep{zhang05}. The central proto-cluster is highly fragmented. We identified 22 condensations with masses ranging from 0.3 to 6.5 M$_{\sun}$. The fragmentation process seems to be determined by thermal pressure in low density regions and by turbulence in the central dense region. The magnetic field may not play an important role in fragmentation.

3. Multiple jet-like high-velocity outflows are revealed in the HCN (3-2) and HCO$^{+}$ (3-2) lines. The orientation and spatial distribution of these high velocity outflows are consistent with previously reported CO outflow. The HCN (3-2) line emission traces more compact and narrower jet-like outflows when compared with HCO$^{+}$ (3-2).

4. We discovered an extremely wide-angle bipolar outflow from low-velocity emission of the HCN (3-2) and HCO$^{+}$ (3-2) lines. The base of this outflow has an opening angle close to 180$\arcdeg$. This outflow seems to be a collection of low-velocity material entrained by the high-velocity outflows due to momentum feedback. The total mass and mass loss rate of the ``EWBO" are 16 M$_{\sun}$ and 6.2$\times10^{-4}$ M$_{\sun}$~yr$^{-1}$, respectively.

5. A hierarchical network of filaments was identified in NH$_{3}$ (1,1) emission. Clear velocity gradients of the order of 10 km~s$^{-1}$~pc$^{-1}$ were found along filaments. The velocity gradients suggest gas inflow along the filaments. The total accretion rate along filaments is $\sim2.1\times10^{-3}$ M$_{\sun}$yr$^{-1}$, which is comparable to the total mass loss rate ($\sim1.7\times10^{-3}$ M$_{\sun}$~yr$^{-1}$) of jet-like outflows and the ``EWBO". The sum of the accretion rate along filaments and mass infall rate along the line of sight is $\sim$3.1$\times10^{-3}$ M$_{\sun}$~yr$^{-1}$, which overcomes the total mass loss rate, indicating that the central cluster is probably still gaining mass.

\section*{Acknowledgment}
\begin{acknowledgements}
We are grateful to the SMA staff. We are grateful to the KVN staff. The KVN is a facility operated by the Korea Astronomy and Space Science Institute. Tie Liu is supported by KASI fellowship. Y. Wu is partly supported by the China Ministry of Science and
Technology under State Key Development Program for Basic Research (No.2012CB821800), the grants of NSFC No.11373009 and No.11433008. S.-L. Qin is partly supported by NSFC under grant Nos. 11373026, 11433004, by Top Talents Program of Yunnan Province. Ke Wang acknowledges support from the ESO fellowship. C.W.L. was supported by Basic Science Research Program though the National Research Foundation of Korea (NRF) funded by the Ministry of Education, Science, and Technology (NRF-2013R1A1A2A10005125) and also by the global research collaboration of Korea Research Council of Fundamental Science \& Technology (KRCF). This work was carried out in part at the Jet Propulsion Laboratory, operated for NASA by the California Institute of Technology. J.-E.L. was supported by the
Basic Science Research Program through the National
Research Foundation of Korea (NRF) (grant No. NRF-
2015R1A2A2A01004769) and the Korea Astronomy and
Space Science Institute under the R\&D program (Project
No. 20151-32018) supervised by the Ministry of Science,
ICT, and Future Planning.  Di Li acknowledges support
from the Guizhou Scientifi c Collaboration Program
(\# 20130421) .

\end{acknowledgements}

\end{document}